\documentstyle[twoside,fleqn,espcrc2,epsfig]{article}
\input epsf.tex
\newcommand{\beq}{\begin{equation}}
\newcommand{\eeq}{\end{equation}}
\newcommand{\beqs}{\begin{eqnarray}}
\newcommand{\eeqs}{\end{eqnarray}}
\newcommand{\laa}{\lambda_{IIA}}
\newcommand{\lop}{\lambda_{I'}}

\newcommand{\da}{{\dot a}}

\title{\vbox{\hbox{\rightline{\rm\small UUITP-21/97}}
\hbox{Creation of Strings in D-particle Quantum Mechanics
\thanks{Talk given by U.D. at Strings 97.}}}}

\author{Ulf Danielsson\address{Institutionen f\"{o}r teoretisk fysik \\
Box 803\\
S-751 08  Uppsala \\
Sweden}
and Gabriele Ferretti\thanks{
Address of G.F. after Sept. 1997: S.I.S.S.A., Strada Costiera 11,
I-34100 Trieste, Italy}
}

\begin{document}

\begin{abstract}
D-particle quantum mechanics in a type I' background is reviewed.
It is also discussed how a string is created when a D-particle is taken
through a D8-brane. The process is found to be dual to the creation of
a D3-brane when a NS5 and D5 brane are passed through each other.
\end{abstract}

\maketitle

\section{Introduction}

The M(atrix) model is a surprising approach to the study of strongly
coupled
type IIA string theory and M-theory \cite{BFSS}. Less explored 
is the corresponding
approach for theories with half the number of supersymmetries. The
heterotic
string M(atrix) model has been developed and studied in e.g. 
\cite{DF,KS,KR,L1,BSS,L2,Rey} but
the number of non-trivial checks that have been performed are much less
than
its successful predecessor.

We will briefly review the quantum mechanics of the type I'
D-particle as presented in \cite{DF}, keeping in mind that 
this is what is also relevant for the heterotic
string. We will furthermore discuss an interesting effect of string
creation,
\cite{BDG,DFK,BGL} and
its duality with a phenomenon discovered by Hanany and Witten \cite{HW}.

The talk is based on the work \cite{DF} and on
the work \cite{DFK} together with I. Klebanov.

\section{The Heterotic Life of the D-particle}

The type IIA D-particle quantum mechanics is obtained by dimensional
reduction
of $N=1$, $D=10$ supersymmetric $SU(N)$ Yang-Mills theory
\cite{DFS,KP,DKPS}. (This system was also studied for different reasons
in \cite{CH,WHN}.) The Hamiltonian is given by
\begin{eqnarray}
        H =  \frac{\laa}{2} E_i^{a2} - {1\over 2}  i f^{abc}
             A_i^a \psi^b\gamma_i\psi^c \nonumber \\
          + {1\over {4\laa}} \left(f^{abc} A_i^b A_j^c\right)^2,
                    \label{IIAhamiltonian}
\end{eqnarray}
where $E_i$ is the conjugate momentum of $A_i$ and $\psi$ is a 16
component
real spinor. All fields are in the adjoint
representation of $SU(N)$ and the number of supersymmetries is 16. This
system 
has been extensively studied and used as a starting point for the 
conjectured M(atrix) description of M-theory.

Another quantum mechanical system of great interest is obtained if we
consider 
type I' string theory, i.e. the T-dual of type I. From the type IIA point
of
view such a system contains two 8-orientifolds ($\Omega 8$), and 16
D8-branes.
For simplicity we will consider a situation where the distance between the
two
$\Omega 8$ is large and 8 of the D8-branes are in the vicinity of each
orientifold. Let us consider a system of $N$ D-particles near one of the
orientifolds.

The quantum mechanics of the system is given by the following Hamiltonian:
\begin{eqnarray}
       & H = \mathrm{Tr}\Bigg\{ \lop \bigg(
        \frac{1}{2} P_i^2  - \frac{1}{2}  E_9^2  \bigg) \nonumber \\
        &  + \frac{1}{\lop}\bigg( \frac{1}{2}[A_9, X_i]^2 -
        \frac{1}{4}[X_i, X_j]^2 \bigg)     \nonumber \\
       &  + \frac{i}{2} \bigg( -S_a[A_9, S_a]  - S_\da[A_9, S_\da] +
        2X_i \sigma^i_{a\da} \{S_a, S_\da\} \nonumber \\ 
        & + \sum _{i=1}^{16} \chi_i^I A_{9IJ}\chi _i^J\bigg) 
         \Bigg\}, \label{hamiltonian}
\end{eqnarray}
where the gauge group is $SO(2N)$.
Here $A_9$ and its conjugate momentum $E_9$ are in the adjoint
representation,
while $X_i$ and its conjugate momenta $P_i$ are in the traceless symmetric 
representation. $A_9$ gives the distance between the
D-particles and their mirror-images, while $X_i$ gives the relative
distances
parallel to the orientifold. Turning to the fermions, $S_a$ is in the
adjoint,
$S_{\dot{a}}$ is in the traceless symmetric and, finally, $\chi _i$ is
in the fundamental representation.

Let us briefly recall why this is the case. The orientifold projection 
$\Omega R$ does two things: through $R$ it performs a reflection in space 
and through $\Omega$ it reverses the orientation of a string. The figure
below
shows what happens.

\begin{center}
\leavevmode
\epsfysize=3cm
\epsfbox{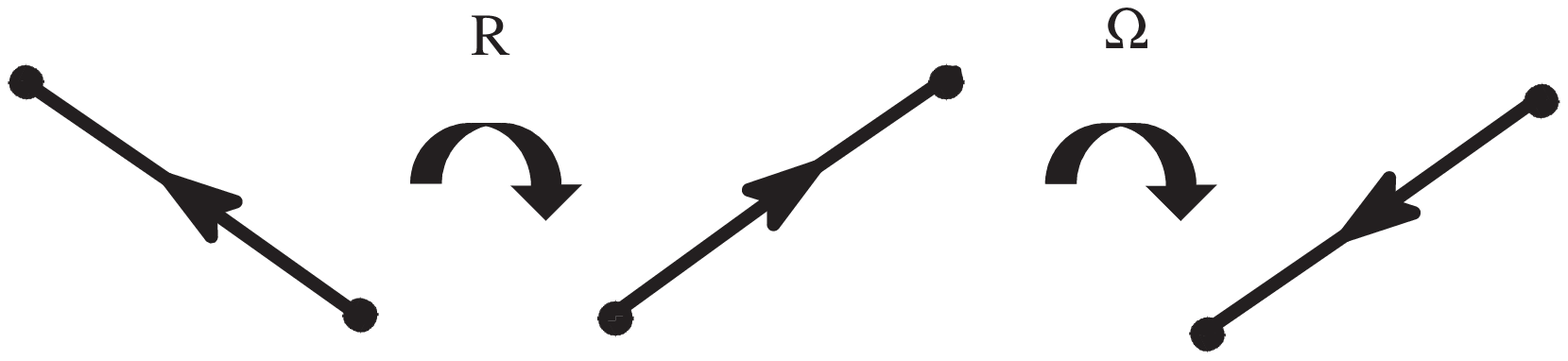}
\end{center}
{\small Fig. 1. The effect of $\Omega$R.}

\vskip 1cm

In case of the field $A_9$ (where '9' refers to the transverse direction) 
it also gives an additional minus sign. This will kill the contribution 
from strings
connecting a D-particle with its own mirror-image, since these strings are
mapped to minus themselves. A simple example is $N=2$,
i.e. two D-particles with their two mirror-images. Excluding the strings
connecting a D-particle with its own mirror-image we find four different
strings not counting orientation. The projection also instructs us to take 
only one particular linear combination with respect to the orientations.
Adding the two strings corresponding to the two independent positions,
i.e.
Cartan elements coming from strings starting and ending on the same
D-particle,
we therefore find $2+4=6$ strings which is the dimension 
of the adjoint of $SO(4)$.

$X_i$ does not have the extra minus sign and the strings joining a
D-particle
with its own mirror-image are therefore kept. In the example that we are 
discussing, this results in four extra strings (counting orientation), in
total
10. This is the dimension of the symmetric representation of $SO(4)$.
Taking
out the trace, which just corresponds to a translation along the
orientifold,
we find the dimension to be 9.

The fermions $S_a$ and $S_{\dot{a}}$ 
have an additional $\Gamma _9$ in the orientifold projection.
Since
\begin{eqnarray}
\Gamma _9 S_a &=& - S_a \nonumber \\
\Gamma _9 S_{\dot{a}} &=& S_{\dot{a}}
\end{eqnarray}
we find that $S_a$ must be in the adjoint and $S_{\dot{a}}$ in the
traceless symmetric.

Finally, we note that the fermions $\chi$ correspond to strings going
between the D-particles and the D8-branes. These, therefore, must be in the
fundamental representation.

\section{... and the Born-Oppenheimer Approximation}

A clear space-time interpretation is obtained when we use the
Born-Oppenheimer approximation, where we integrate out the fast modes and
only the slow ones (corresponding to Cartan-elements of the algebra)
remain.
In order to do this we turn on $A_9$ to separate the D-particles from the
orientifold and $X_i$ to separate them from each other. The various fields
decompose into slow and fast as follows:
\begin{eqnarray}
\hbox{dim}(\hbox{adj}_{SO(2N)}) +
8\hbox{dim}(\hbox{sym}_{SO(2N)}) \nonumber \\
=N(2N-1)+8(N(2N+1)-1) \nonumber \\
=\underbrace{2N^2+7N-8}_{\hbox{slow}} + \underbrace{16N^2}_{\hbox{fast}}
\end{eqnarray}
Modding out the gauge equivalent slow modes, whose number is
\beq
\hbox{dim}(SO(2N))-\hbox{rank}(SO(2N))=2N^2-2N
\eeq
one sees that the dimension of the moduli space is $N+8(N-1)$. 
$N$ is the number of
neutral elements in the adjoint while $N-1$ is the number of neutral
elements
in the traceless symmetric. Let us now consider the special case $N=1$. We
find
one slow mode, i.e. the distance between the single D-particle and the
orientifold, and 16 fast ones. There is also 8 heavy fermions,
$S_{\dot{a}}$,
in the symmetric representation giving 16 real fermions.

It is important to note that this is half the number of fermions as
compared
to type IIA quantum mechanics.

Let us now quantize the theory using the Born-Oppenheimer approximation. 
If we consider the contribution from the orientifold $\Omega 8$,
we find the following effective potential for the slow mode:
\beq
V_{\Omega 8}(r) = 2r\sum_{i=1}^{16}(N_i^B +1/2) +
    2r\sum_{i=1}^{8}(N_i^F -1/2).
\eeq
Here we have used the 16 real fermions to construct 8 raising and 8
lowering
operators. We see that the ground state energy do not seem to cancel!
However, we also need to take the D8-branes into account. We have $8+8=16$
fermions $\chi _i$ in the fundamental representation. 8 coming from
strings
connecting the D-particle and the D8-branes, and the same number from
strings
connecting the D-particle with the mirror-images of the D8-branes. In
total
we have 32 real fermions. Since we above found that we had half the
required
number of fermions, it now seems as if we are overshooting. However the 
coupling between the two types of fermions and the field $A_9$ is
different
due to the different representations. This supplies the needed factor one
half.
This can also be understood from the fact that in one case we are dealing
with 
the distance to the D-particle mirror-image, in the other case with the
distance to a D8-brane at the orientifold. The contribution to the
potential
coming from the D8-branes is thus:
\beq
V_{D8}(r) = r\sum_{i=1}^{16} (N_i^f -1/2),\label{poteight}
\eeq
and the total potential is hence given by:
\beq
V(r) =  V_{\Omega 8}(r) + V_{D8}(r).\label{pot}
\eeq
We now see that the ground state energy cancels, implying that the 
D-particle at rest does not feel any force. Furthermore we observe that a
D8-brane seems to be repulsive, while a 8-orientifold seems to be
attractive.

\section{Taking a D-particle through a D8-brane}

We can now generalize the above potential slightly by allowing the
D8-branes
to be positioned off the 8-orientifold. We then find that the 
potential (\ref{poteight}) generalizes to
\beqs
V_{D8}(r)=\sum_{i=1}^{8} \bigg(|r-m_i |(N_i^{fR}
-1/2)\nonumber\\
+|r+m_i| (N_i^{fL} -1/2)\bigg). \label{nypot}
\eeqs
We will now ask the following question: what happens if a 
D-particle is taken 
through one of the D8-branes? 
Looking at the potential we seem to conclude
that there now is a net force on the D-particle, trying to push it towards
the orientifold. This is due to the change of sign of the force from the
D8-brane that the D-particle crossed. But if a string is created, i.e. 
$N_8^{fR}$ jumps to $N_8^{fR}=1$, the force is canceled! Now, does this
really happen?

\section{A U-dual Situation}

We will now show that a string is indeed created by using U-duality.
Hanany
and Witten have showed that a D3-brane is created when a D5-brane,
spanning directions 1-2-6-7-8, is passed through a NS5-brane spanning 
directions 1-2-3-4-5. The resulting D3-brane will span the 
directions 1-2 along which the
two five-branes intersect and the transverse direction 9.

This result can now be used in the situation that we are interested in.
First
we T-dualize along directions 1 and 2. The NS5-brane remains an NS5-brane,
while
the D5-brane becomes a D3-brane along 6-7-8 and the D3-brane becomes a
D-string
along 9. Then we use S-duality of type IIB string theory which maps the 
NS5-brane to a D5-brane. Similarly the the D-string becomes a fundamental 
string. The D3-brane remains a D3-brane. The final step involves
T-dualities
along the directions 6, 7 and 8. The D5-brane now becomes a D8-brane and
the
D3-brane becomes a D-particle. The fundamental string remains a
fundamental
string. This is precisely the setup that we are after. We conclude that
a fundamental string must be created when a D-particle passes through a 
D8-brane.

Instead of using S-duality we might view everything from a M-theory
perspective, as in the diagram below. We see how two different
compactifications (on direction 10 and direction 1 respectively)
followed up with T-dualities give on the one hand the Hanany-Witten
result,
and on the other hand the result that we are after.

\begin{center}
\leavevmode
\epsfysize=5cm
\epsfbox{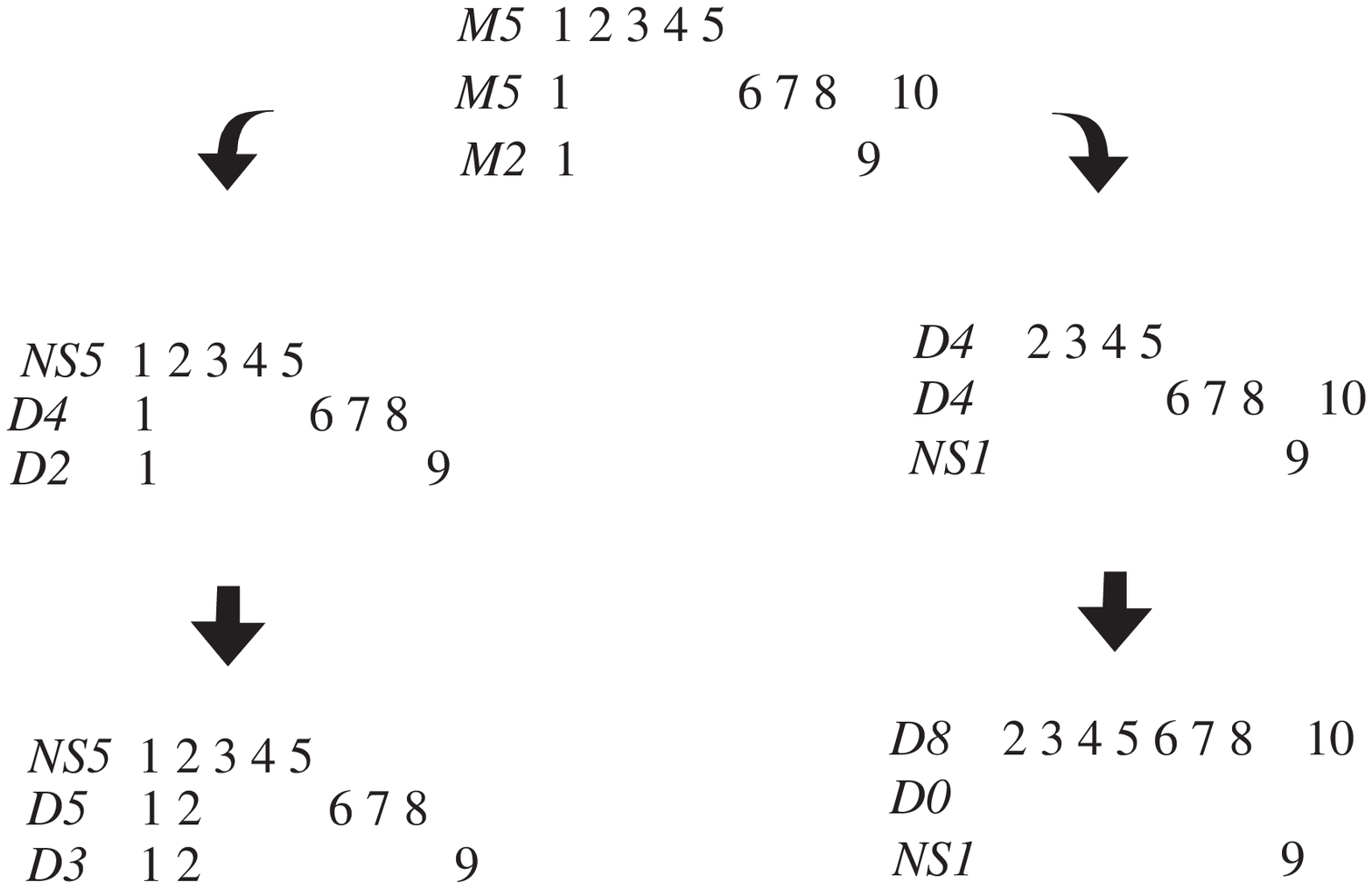}
\end{center}
{\small Fig. 2. Creation through duality.}

\vskip 1cm

\section{Putting it All Together}

The previous results were all written in a gauge where $A_0=0$. In general
it is found that not only an induced potential, but also an induced
Chern-Simons
term is generated. For instance, a D8-brane gives rise to, \cite{BSS},
\beq
-\frac{1}{2}\hbox{sign}(A_9-m)(A_9-m+A_0),
\eeq
generalizing the zero-point energy part of (\ref{nypot}) if $A_9 =r$.
But this is not all, in addition to the induced terms
we might also have bare terms. We will now discuss how to interpret these
bare and induced terms from a stringy point of view.

We begin by discussing the force between two parallel $p$ and $q$
branes 
at rest, \cite{polch}.
If $p-q=0$ the contributions are schematically
\beq
NS+R+NS(-1)^F=0
\eeq
from the open string point of view. From the closed string point of view
the
first two terms correspond to an NSNS-exchange of gravitons and dilatons
giving an
attraction, the last term is due to RR-exchange and gives a repulsion in
such
a way that the total force cancels. 

When $p-q=4$ we have
\beq
NS+R=0.
\eeq
{}From the closed string point of view we have a NSNS-exchange only,
with the gravitons and dilatons canceling all by themselves.

When $p-q=8$, which is the situation that we are interested in, we have, 
\cite{Lif},
\beq
NS+R+R(-1)^F=0.
\eeq
The first two terms, due to graviton and dilaton exchange, gives a
repulsion that is canceled by the third term. 

The question now arises how
to interpret the third term from the closed string perspective.
For an isolated D-particle--D8-brane system in type IIA string theory
there is an argument that interprets the $R(-1)^F$ term as due
to the creation of ``half'' a string between the branes. It would be 
interesting to put this on more rigorous grounds; however, for the case of
real 
interest to us, that is the type I' system, the situation is much 
better established: when the D-particle is in the ``symmetric'' position
(with 8 D8-branes and one 8-orientifold on each side) all the forces
cancel
out and as it moves across a D8-brane a string is created to cancel the
unbalanced force. Thus, due to string creation, the D-particle feels no
force everywhere. We shall review these arguments below.
  
In the type IIA scenario, one can heuristically argue in the following
way;
the D8-brane is the source of a 10-form
field strength with dual $F= ^* F_{10}$ with the following coupling to the
D-particle:
\beq
\mu _{0} \int d\tau F A_{0}.
\eeq
This implies that it will
appear like an electric charge on the D-particle world line. Since this is
also the way that the endpoint of a fundamental string would appear, we 
conclude that a fundamental string must end on the D-particle. This is
what gives rise to the extra $R(-1)^F$ force. One should note, however,
that the term corresponds to a tension equivalent to ``half'' a string.
As discussed in e.g. \cite{PW}, $F$ is piece-wise constant so that
when the D-particle goes through the D8-brane $F$ jumps by $\mu_{8}$, and
therefore $\mu _{0} F$ jumps by $\mu _{0} \mu _{8} = \frac{1}{2 \pi \alpha
'}$. We are working in units where $\frac{1}{2 \pi \alpha
'}=1$.
This can be interpreted as the creation of a full string that cancels
the original half a string with orientation, say, towards the D-particle
and gives half a string with orientation away from the D-particle.

The type I' setting luckily allows for
an interpretation without half-strings. In that case, by T-dualizing
a type I configuration without Wilson lines, one obtains the symmetric
configuration described above with  $F=0$ in the region where the
D-particle
has 8 D8-branes on each side. The jump in
$F$ felt by the D-particle as it moves through some of the D8-branes
corresponds to the creation of a string that we have discussed. We believe
that the creation of the string, with consequent cancellation of the net
force
is also necessary if one has to make sense of the construction of
\cite{KS}
where one obtains the extra components of the $E_8$ vector multiplets as
bound states at threshold; a non zero potential would in fact lower the 
energy of such states below the continuum.

To summarize, we conclude that the NSNS repulsion 
corresponds to the
{\it induced} term in the D-particle quantum mechanics. Recall that
we found a D8-brane to be repulsive. The term $R(-1)^F$ that we have
identified as coming from $F$  
must therefore correspond to the {\it bare} term.
Looking at the potential we see that the induced terms correspond to
the zero-point energy while the bare terms should be associated with
excited 
states, i.e. real strings.
 
The string theoretical annulus calculation seems to 
give a result that do not include
the string-creation effect. Without this effect, the potential is
asymmetric with a force on one side and no force on the other, \cite{BGL}.
It should be noted that in the type I' case, the contribution from
$R(-1)^F$
actually cancels everywhere (excluding string creattion effects). The reason why the force seems to jump when
a D8-brane is passed is hence simply related to the change of 
direction of the well
understood graviton/dilaton force. The string creation is an effect
that
have to be added and, indeed, cancels the jump.
\footnote{In \cite{HoWu} these effects are explained from a M(atrix)-model
point of
view using 5-branes and membranes. Through T-duality the authors conclude
that in the case of one D8-brane, the potential of \cite{BGL} is
reproduced. 
However, the authors also show how the
adiabatic passage of a D-particle through the D8-brane creates a string.
This cancels the force.}

\section*{Acknowledgments}

We would like to thank  Igor Klebanov 
for a fruitful collaboration and Adam Schwimmer for discussions.



\end{document}